\documentclass[aps,prl,twocolumn,showpacs,groupedaddress]{revtex4-1}
\usepackage{graphicx}
\usepackage{epstopdf}
\begin{document}

\title{Theory of dipolaron solutions}
\author{Dima Bolmatov$^{1}$}
\thanks{d.bolmatov@gmail.com}
\author{S. Bastrukov$^{2}$}
\author{P.-Y. Lai$^{3}$}
\author{I. Molodtsova$^{4}$}
\address{$^1$ Baker Laboratory, Cornell University, Ithaca, New York 14853-1301, USA}
\address{$^2$ Data Storage Institute, A*STAR (Agency for Science, Technology and Research), Singapore 117608, Singapore}
\address{$^3$ Department of Physics and Center for Complex Systems, National Central University, Chungli, Taiwan 320, R.O.C.}
\address{$^4$ Joint Institute for Nuclear Research, Dubna 141980, Russia}
\begin{abstract}
A fundamental task of statistical physics is to predict the system's statistical properties and compare them with observable data. We formulate the theory of dipolaron solutions and analyze the screening effects for permanent and field-induced dipolarons. The mathematical treatment of the collective behaviour and microscopical morphology of dipolaron solutions are discussed. The presented computations show that the electric field shielding of dipolarons in dielectric nanosolutions is quite different from that of counterionic nano-complexes of Debye-H\"uckel theory of electrolytes. The limiting case of screening length $\lambda=0$ in dipolaron solutions corresponds to Coulomb law for the potential and field of uniformly charged sphere. 
\end{abstract}


\maketitle
$Introduction$. 
A molecular description of matter using statistical mechanical theories and computer simulation is the key to understanding and predicting the properties of liquids and fluids. The analytical, numerical, and experimental study of properties of fluids is a well-established discipline of statistical mechanics. The theory of ionic solutions has been one of the most important and fundamental problems in statistical physics throughout the last century. There is general agreement that the Debye-H\"uckel (DH) theory was a revolution in the understanding of the properties of homogeneous ionic solutions (electrolytes). The theory of electrolyte solutions has been the object of a huge number of scientific results during the 20th century and last decade \cite{alex-1, koe-1,santos-1}, due to the great amount of applications in the most diverse areas of basic and applied research \cite{tad-1, jon-1, netz-1, denton-1}. Progress in this field has been possible predominantly because of an adequate knowledge of the interionic interactions. The formalisms of electrostatics, statistical mechanics and hydrodynamics have been successfully applied to situations where the long-range Coulombic interactions prevail over solvent-solvent, ion-solvent or short-ranged ion-ion forces. 

One of the earliest efforts to evaluate the ionic distribution functions was undertaken by Debye and H\"uckel in their classical paper of 1923 \cite{debye}. Their results were extremely influential, mainly because more elaborate liquid state theories were not developed until the 1960s and 1970s and a common approach to electrolytes and non-electrolytes was not possible. DH results are now recognized as the universally valid limiting law for ionic thermodynamic quantities at infinite dilution. The importance of DH formalism is still enormous nowadays, when it continues to be the theoretical basis of most practical applications, particularly in physics of liquids, solid-state plasmas, astrophysics and nuclear reactor physics. 

Debye and H\"uckel introduced a model of ionic solutions where the ions are treated as ionic point charges which interact by means of a Coulomb potential in a uniform dielectric background. They assumed that the ions are distributed according to an exponential distribution law (Boltzmann distribution) characteristic of a system in
thermal equilibrium with a heat reservoir (the solvent). The key concept in DH theory is the ionic atmosphere, a spatial separation of charge made up of mobile ions which balance the charge of the central ion, that allows the understanding of the particular ordering inside the bulk solution. The spatial range of correlations itself is determined by the size of this charge inhomogeneity. 


A great number of attempts to go beyond the Debye-H\"uckel theory have been reported, based on the overcoming of the linearization of the Poisson-Boltzmann equation \cite{gronwall}, the introduction of ionic pairing \cite{bjerrum}, or the existence of a pseudoreticular structure in the dense regime of an ionic solution \cite{bahe}. Integral equation techniques have also been used for the obtaining of the pair correlation functions from the Ornstein-Zernike equation: the mean spherical approximation \cite{waisman}, its thermodynamically consistent generalization \cite{lebowitz}, or their improvements based on cluster resummation techniques, the optimized random phase approximation \cite{chandler}. Other integral equation techniques have been tested
with success, such as the Percus-Yevick type equation \cite{allnatt}, the hypernetted chain equation \cite{friedman}, and path integral approaches \cite{efimov,bolmatov1}. Varela and co-authors reviewed the theoretical and numerical results reported throughout the last century for homogeneous and inhomogeneous ionic systems \cite{varela1}. 

\begin{figure}
\begin{center}
{\scalebox{0.3}{\includegraphics{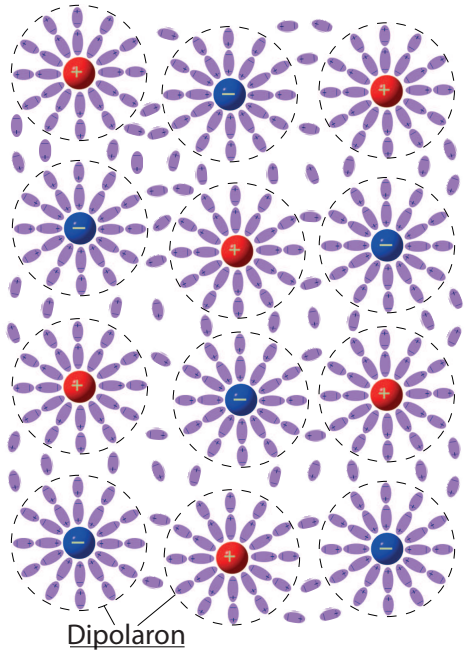}}}
\end{center}
\caption{A cloud of radially polarized molecular dipoles and monopole charges of counterions form nanostructured units-{\it dipolarons}. The microscopical morphology of dielectric solution is determined by and derived from Eqs.\ref{field},\ref{field1}.}
\label{fig1}
\end{figure}

This work is consistent with the ongoing effort in elucidating and understanding the structure and properties of disordered matter such as liquids \cite{bolnat,widom1,grim,widom2,monaco,hosokawa, guarini}, glasses \cite{ruocco,aaron} and colloids \cite{boh,yuf}. Inspired by recent progress in the liquid state of matter \cite{bolsym,bol2,bol3,bol4} we formulate the theory of dipolaron solutions. The mathematical treatment of the collective behaviour and microscopical morphology of dipolaron solutions (see Fig.\ref{fig1}) are the chief problems to be discussed in this paper. {\it We postulate the dipolaron structure; dipolaron is a nano-complex formed by finite size ions and polar solvent molecules surrounding these ions}. Molecular dipole moments may cooperatively enhance or counteract existing entropic depletion and electrostatic forces. We suggest to find the functional form of electric field and then from the first Maxwell equation to find the screened potential.  A complete dielectric homogeneity across the system is implied.

$Basic$ $equations$ $of$  $dipolaron$ $electrostatics$. The electrostatic state of an isotropic dielectric continuous medium is described by equations
(e.g., \cite{smyth})
\begin{eqnarray}
{\bf D}=\epsilon_{0}{\bf E}+{\bf P}
\label{eqn1} 
\\
\nabla\cdot{\bf E}=\rho
\label{eqn2}
\end{eqnarray}
where ${\bf D}$ is the field of electric displacement, ${\bf P}$ is the density of electric polarization - electric
dipole moment per unit volume and $\epsilon_{0}$ stands for the electric permittivity of free space. The Eq.(\ref{eqn2}) is the Maxwell equation for the electrostatic field ${\bf E}$ (differential form of
Coulomb law). $\rho$ is the total density of electric charge, which is a sum of the free charge
density $\rho_{f}$ and bound charge density $\rho_{b}$ and can be defined as: 
\begin{eqnarray}
& & \rho=\rho_{f}+\rho_{b} \\
& & \nabla\cdot{\bf D}=\rho_{f} \\
& & \nabla\cdot{\bf P}=-\rho_{b}
\end{eqnarray}
As we mentioned before, the main difference between dielectric fluids and electrolytes is that
the former can be usually thought of as a non-conducting suspension composed of bound
charges dissolved in a fluid matrix of electrically polarized molecules, whereas electrolytes are
an electrically conducting solution (of second type) whose conductivity is attributed to the
mobility of ions, regarded as free (non-bounded) charges. Taking this into account
and putting $\nabla\cdot{\bf D}=\rho_{f}=0$ one has 
\begin{equation}
\nabla\cdot(\epsilon_{0}{\bf E})=\rho_{b}
\end{equation}
Bound charges of dipoles set up electric dipoles in response to an applied electric field ${\bf E}$ and polarize other nearby dipoles causing to line them up. The net accumulation of charge from the orientation of the dipoles is the bound charge. he response of the rest of dipolar solution to the electric field of bound charges is determined by dipolarons-nanostructures, composed of monopole charges of counterions surrounded by a cloud of radially polarized molecular dipoles, forming dipolar atmospheres. The macroscopic bound charge density of dielectric solution can be reperesented as $\rho_{b}=q_{i}\tilde{n}_{i}/\epsilon$. Index $i$ labels counterions carrying the charge $q_{i}=ez_{i}$ and $\tilde{n}_{i}$ is the effective particle density of bound charges associated with the ion of $i$-th type which takes into account polarization of molecular dipoles by the field of given ion. We assume the Boltzmann distribution 
law, according to which the average local concentration $\tilde{n}_{i}$ is
\begin{equation}
\tilde{n}_{i}=n_{i}\exp{\left(-\frac{N_{l}{\bf p}_{l}\cdot{\bf E}}{k_{B}T}\right)}
\label{Boltz}
\end{equation}
where $\epsilon$ is electric permittivity of the solvent, $N_{l}$ is the number of dipole-polarized molecules
of $l$-type with a dipole moment ${\bf p}_{l}$. $n_{i}$ is the effective particle density of ions in the absence of solvent dipole molecules, as in the case of dielectric the exterior field and its potential can be represented in the form of dissolved ions in fluid matrix of electrically passive, unpolarizable molecules (i.e., molecules which are non-polar and insensitive to fields of ions). Eq.(\ref{Boltz}) is justified by the fact that the average particle density of bound charges of dielectric solution is in thermal equilibrium at elevated temperatures. Considering this we obtain a non-linear equation of dipolaron electrostatics
\begin{equation}
\nabla\cdot{\bf E}= \frac{q_{i}n_{i}}{\epsilon_{0}\epsilon}\exp{\left(-\frac{N_{l}{\bf p}_{l}\cdot{\bf E}}{k_{B}T}\right)}
\end{equation}
Expanding Eq.(\ref{Boltz}) in the form: $e^{-x}=1-x+\frac{x^2}{2!}-\frac{x^3}{3!}+\ldots$ and neglecting the terms of higher order (the second and higher order terms) of expansion we obtain the following expression
\begin{equation}
\nabla\cdot{\bf E}+\frac{q_{i}n_{i}}{\epsilon_{0}\epsilon}\frac{N_{l}}{k_{B}T}{\bf p}_{l}\cdot{\bf E}=\frac{q_{i}n_{i}}{\epsilon_{0}\epsilon}
\label{maineq}
\end{equation}
This equation constitutes the mathematical basis for the further analysis of the electric
field shielding in the dielectric solutions.

$Shielding$ $of$ $dipolaronic$ $electrostatic$ $fields$. In view of general results derived in the preceding section, we focus on application of Eq.(\ref{maineq})
to electrostatics of in-medium dipolarons with a spherical shape of radius $R$. In accord with the standard approach to the problem of electrostatic fields inside and outside a sphere where the interior ${\bf E}_{in}={\bf E}(r<R)$ and exterior ${\bf E}_{e}={\bf E}(r>R)$ fields should be computed from the two equations
\begin{eqnarray}
& & \nabla\cdot{\bf E}_{in}+\eta_{l}{\bf p}_{l}\cdot{\bf E}_{in}=\frac{q_{i}n_{i}}{\epsilon_{0}\epsilon}, \ \  r\leq R  \\
& &\nabla\cdot{\bf E}_{e}+\eta_{l}{\bf p}_{l}\cdot{\bf E}_{e}=0, \ \  r> R
\end{eqnarray}
supplemented by boundary conditions on the surface of dipolaron of radius $R$, where $\eta_{l}=\frac{q_{i}n_{i}}{\epsilon_{0}\epsilon}\frac{N_{l}}{k_{B}T}$. $\eta_{l}$ is the in-medium parameter which accounts for the dipole polarization of solvent molecules from dipolar atmosphere by the field of monopole charge of central micro-ion. In the current manuscript we are going to consider two kinds of dipolarons 
distinquished by the type of solvent molecules forming its dipolar atmosphere. The first model belongs to aqua- and aqua-like solutions composed of polar molecules of solvent, that is molecules endowed with
permanent dipole moments of constant magnitude, ${\bf p}_{l}=$constant. The nano-complex of this type is referred to as a $permanent$ $dipolaron$. The second model belongs to a case where dipolar atmosphere consists of $N_{l}$ non-polar molecules whose dipole electric moments ${\bf p}_{l}$ are induced by local fields ${\bf E}$ of macro-ions:
${\bf p}_{l}=\alpha_{l}{\bf E}$, where constant parameter $\alpha_{l}$ is the molecular polarizability of solvent molecule of $l$-sort. The in-medium dipolaron of this type is referred to as a $field$-$induced$ $dipolaron$. 
Since our prime goal is the in-medium effect of electric field shielding, we focus on exterior fields of  dipolarons whose central symmetric shape suggests that the local fields have only one radial component
\begin{equation}
{\bf E}({\bf r})=-\nabla\Phi({\bf r}), \ \ [E_{r}(r)=-\nabla_{r}\Phi(r), \ E_{\theta}=0, \ E_{\phi}=0]
\end{equation}

{\it Exterior field of permanent dipolaron}. Taking into account the last remark presuming that $N_{l}{\bf p}_{l}{\bf E}_{e}=N_{l}p_{l}E_{e}$, one finds that the radial component of exterior field of permanent dipolaron obeys the liner equation
\begin{equation}
\nabla_{r}E_{e}(r)+\lambda^{-1}E_{e}(r)=0, \ r>R, \ \lambda^{-1}=\frac{q_{i}n_{i}}{\epsilon_{0}\epsilon}\frac{N_{l}p_{l}}{k_{B}T}
\end{equation}

\begin{figure}
\begin{center}
{\scalebox{0.55}{\includegraphics{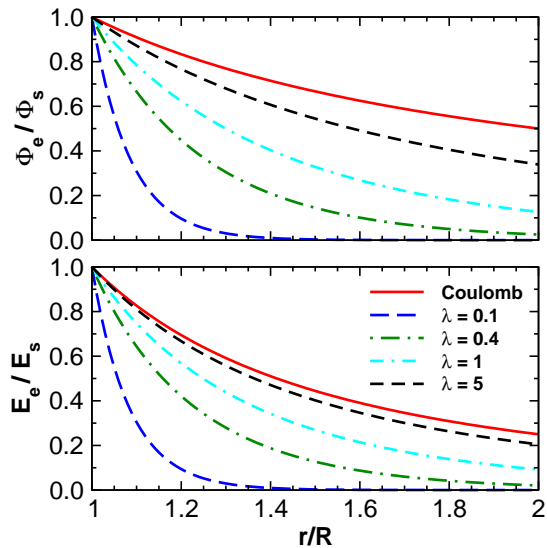}}}
\end{center}
\caption{Normalized to their surface values the external potential $\Phi_{e}/\Phi_{s}$ and the radial
component of external field $E_{e}/E_{s}$ of permanent dipolaron computed as functions of $r/R$
for a different screening length values $\lambda$. The case $\lambda=0$ corresponds to the exterior field of uniformly charged sphere in dielectric fluid composed of electrically passive molecules, insensitive to
dipolar polarization.}
\label{kar}
\end{figure}
With the help of substitution $E_{e}(r)=s(r)/r^2$ we obtain the equation for $s(r)$ permitting the
general solution 
\begin{equation}
\frac{ds}{dr}+\lambda^{-1}s=0 \ \longrightarrow s=s_{0}e^{-r/\lambda}
\end{equation}
For the radial component of external electrostatic field of permanent dipolaron we gain
\begin{equation}\label{field}
E_{e}(r)=\frac{q}{4\pi\epsilon_{0}\epsilon}\frac{e^{-r/\lambda}}{r^2}
\end{equation}
$q$ is the total charge of dipolaron and its density accounts for the internal polarization of polar molecules due to fields of central ion. The potential of this field
\begin{eqnarray}
\frac{d\Phi_{e}}{dr} &=& -\frac{q}{4\pi\epsilon_{0}\epsilon}\frac{e^{-r/\lambda}}{r^2}\\
\Phi_{e}(r) &=& -\frac{q}{4\pi\epsilon_{0}\epsilon}\left[\frac{Ei(1,r/\lambda)}{\lambda}-\frac{e^{-r/\lambda}}{r}\right]
\end{eqnarray}
where $Ei(n,x)=\int_{1}^{\infty}\frac{e^{-xt}}{t^n}dt$ is the Exponential integral (e.g.,\cite{stegan}). 
In Eq.\ref{field} the exponential factor exhibits the in-medium effect of electric field shielding. This effect is controlled by parameter $\lambda$ and depends on characteristics of both counterions and molecules of solvent: the
larger $\lambda$ the smaller effect of shielding. The magnitude of the screening length $\lambda$ bears information about all constituents of dielectric nanosolution as it depends upon bound-charged density of all counterions $q_{i}n_{i}$, electric permittivity $\epsilon$ and dipole moments of solvent molecules $p_{l}$.
The shielding of the exterior field of permanent dipolaron is demonstrated in Fig.\ref{kar}: we plot exterior field and potential as functions of distance from the dipolaron surface (normalized to its radius).

{\it Exterior field of field-induced dipolaron}. We now consider a model of field-induced dipolaron which can be appropriately applied to dielectric nanosolutions composed of non-polar molecules (easily polarized). Under the action of an ionic field the molecules of dipolar atmospheres acquire a dipole electric moments $\bf{p}_{l}=\alpha_{l}\bf{E}$, where $l$ labels the dipolar type of molecule and the exterior electrostatic field of field-induced dipolaron is described by the non-linear equation
\begin{equation}
\nabla\cdot{\bf E}_{e}+\zeta{\bf E}^{2}_{e}=0, \ {\bf E}_{e}=-\nabla\Phi_{e},\ r>R
\end{equation}
where $\zeta_{l}=\frac{q_{i}n_{i}}{\epsilon_{0}\epsilon}\frac{N_{l}\alpha_{l}}{k_{B}T}$, which can be solved analytically. After tedious analytic computations we present here the final result
\begin{equation}
{\bf E}_{e}=\frac{q}{4\pi\epsilon_{0}\epsilon}\frac{1}{r(r-q\zeta)}{\bf e}_{r}
\end{equation}
which can be easily verified by simple computation. The exterior field and its potential can be represented in the form
\begin{eqnarray}\label{field1}
& & {\bf E}_{e}=\frac{q}{4\pi\epsilon_{0}\epsilon r^2}\left(1-\frac{\Lambda}{r}\right)^{-1}{\bf e}_{r}, \\
& & \Phi_{e}(r)=-\frac{q}{\Lambda}\ln{\left( 1-\frac{\Lambda}{r}\right)}
\end{eqnarray}
where $\Lambda=\frac{q_{i}^{2}n_{i}}{4\pi\epsilon_{0}\epsilon}\frac{N\alpha}{k_{B}T}$, highlighting the effect of the non-Debye field screening. Here $q$ is the total charge of field-induced dipolaron. 
\begin{figure}
\begin{center}
{\scalebox{0.55}{\includegraphics{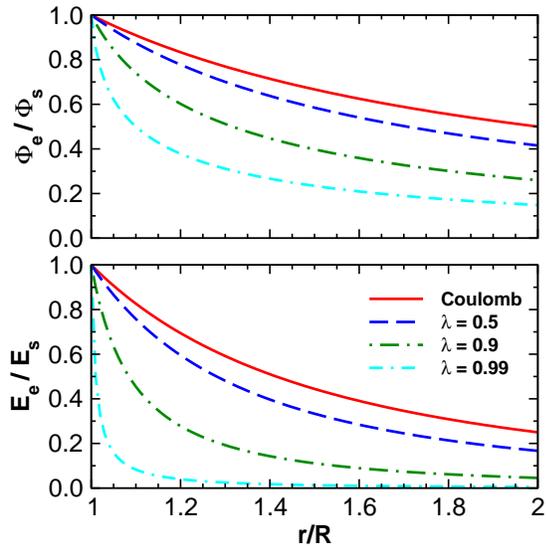}}}
\end{center}
\caption{The potential $\Phi_{e}/\Phi_{s}$ and the field $E_{e}/E_{s}$ (normalized to their 
surface values) as a function of $x=r/R$ of field-induced dipolaron computed for different values
of parameter $\lambda$ regulating intensity of electric field shielding. The limiting case of $\lambda=0$ corresponds to Coulomb law for the potential and field of uniformly charged sphere.}
\label{pic}
\end{figure}
Unlike the exponential shielding of electric field of the permanent dipolaron, the dipolar atmosphere of field-induced dipolaron leads to the electric field screening which is described by $\left( 1-\frac{\Lambda}{r}\right)$
factor. The above characteristics can be conveniently represented via dimensionless parameters of the field shielding $\lambda$ and distance (normalized to the dipolaron radius) $x=r/R>1$ which is measured from the surface of dipolaron: $\lambda=\Lambda/R$, $0<\lambda<1$, the radial component of external field 
\begin{eqnarray}
& & E_{e}=\frac{q}{4\pi\epsilon_{0}\epsilon x^2 R^2}\left(1-\frac{\lambda}{x}\right)^{-1}, \\
& & \Phi_{e}(r)=-\frac{q}{4\pi\epsilon_{0}\epsilon\lambda R}\ln{\left( 1-\frac{\lambda}{x}\right)}
\end{eqnarray}
The case of $x = 1$ corresponds to the surface values of these quantities. In Fig.\ref{pic}, we plot
the radial component of exterior electric field and potential of the field-induced dipolaron
as functions of $x$ computed at different values of shielding length $\lambda$. The smaller $\lambda$ the less in-medium effect of the field shielding. The limiting case of $\lambda=0$ corresponds to Coulomb law for the potential and field of uniformly charged sphere. 

$Conclusions$. To the best of our knowledge some statistical problems of complex fluid thought of as ion-dipole mixture have first been discussed in paper entitled {\it On the theory of dipolar fluids and ion–dipole mixtures} \cite{derek}. In particular, it has been shown to emerge the Debye-H\"uckel form for the asymptotic behavior of the potential of mean force between ions for arbitrary density of the dipolar solvent in the limit of low ionic concentration.

In this work we go further considering dipole-ion cofiguration (see Fig. 1 and dubbed to as {\it dipolarion})
with focus on the problem of the electric field shielding. As it clearly seen from Fig. 1 the dopile-ion agglomerations possesses electric  charge,  contrary to DH constituents and this is the reason why in present the basic equation of in-medium electrostatics has the form different from classical DH equation. Also, at this point it worthwhile to stress that that dipolarion is substantially different from dipolaron (electron-hole constituent) of solid state physics.

The theory introduced in this paper leads to a correct physical picture of non-Debye screening which is experimentally known for a wide class of materials. The introduced equations are interesting in their own right due to their capability of consistent mathematical description and physical treatment of the in-medium effect of electric field shielding which is caused by the polarization of molecular dipoles of
solvent by fields of counterions. This capability has been demonstrated by numerical analysis
of obtained analytic solutions for exterior fields of permanet and field-induced dipolarons.
Calculus of the effective parameters and investigation of the thermodynamic properties of dipolaron solutions is the subject of forthcoming papers.

$Acknowledgements$. The authors are indebted to Wei-Yin Chiang for helpful assistance and Andrey Rogachev for fruitful discussions. This work is partly supported by NSC of Taiwan, under grant nos. 98-2112-M-008-023-MY3, and NCTS of Taiwan. D. Bolmatov thanks Cornell University for support.

\end{document}